# Reconfigurable Strain Gradient Polarity in Crystalline Oxide Nanomembranes for Controlled Bending of Functional Materials

*Tiffany C. Wang*[1,2*], *Minyong Han*[1,2], *Varun Harbola*[1,3†], *Harold Y. Hwang*[1,2*]

1 Stanford Institute for Materials and Energy Sciences, SLAC National Accelerator Laboratory; Menlo Park, CA 94025, United States.

2 Department of Applied Physics, Stanford University; Stanford, CA 94305, United States.

3 Department of Physics, Stanford University; Stanford, CA 94305, United States.

ABSTRACT

We report the fabrication and mechanical characterization of a "bubble" geometry for accessing local strain gradients using freestanding, single-crystalline manganite nanomembranes: a switchable bistable nanodrum, with opposite strain gradient polarities. By leveraging epitaxial strain as a source of pre-strain and with control of geometrical and mechanical boundary conditions, the fabricated device can support strain gradients with strain variation ($\Delta\varepsilon$) ranging from 0.01% to 1%. Switching energetics can be designed to configure the bubble morphology. By providing a mechanical framework for sustained strain gradients, this platform supports scalable oxide membrane applications such as the mechanical manipulation of magnetism, coupled to local probes.

The physical properties of solid-state systems originate from their periodic lattice environment and the atomic interactions therein. Reversible and dynamic strain tuning has become a powerful control parameter for manipulating crystalline materials by changing interatomic distances [1–4]. In oxides, the fabrication of freestanding single-crystalline membranes [5] further enables the application of extreme strain states, expanding possibilities beyond the van der Waals (vdW) materials [6–8]. When released from the substrate and transferred onto an arbitrary elastic platform, membranes manifest surprisingly high stretchability [2,9,10]. Additionally, oxide membranes can accommodate a high degree of bending in a regime which rigid bulk crystals or on-substrate thin films cannot reach. Building upon our previous work demonstrating the extreme elasticity and strain gradient mechanics of freestanding membranes [9,10] this study advances the structural capability by engineering bistable nanodrums. These bubbles form spontaneously when a membrane, leveraging epitaxial strain as a source of "pre-strain," is transferred onto a commercially available porous substrate; and the residual compressive strain is released through out-of-plane buckling into bistable states. This bending, or strain gradient, breaks inversion symmetry, and can activate unique categories of emergent physical phenomena, including polar distortions [11], ferro- and flexo-electricity, and magnetic textures [12–17]. The aim of this work is to establish a mechanical and structural engineering platform to controllably access these states. The ability to enforce symmetry breaking and stable strain gradients in a solid-state membrane presents a highly tunable environment that can be leveraged for material engineering applications such as the mechanical manipulation of magnetism, coupled to local probes.

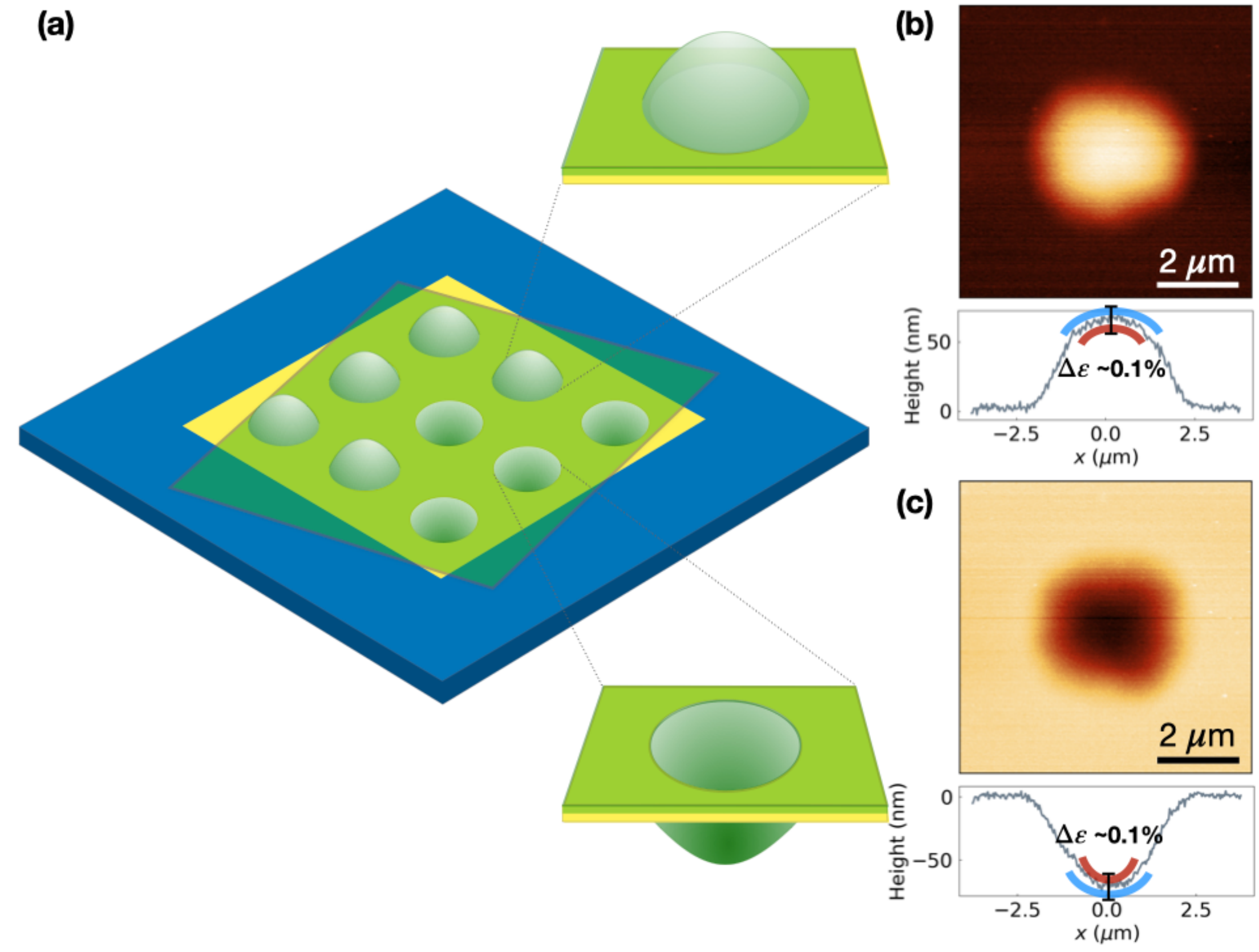

**Figure 1. Illustration of spontaneous bistable bubble formation.** (a) Schematic bistable states of freestanding 30 nm thick $La_{0.7}Sr_{0.3}MnO_3$ (LSMO) membrane (green) upon release and transfer onto commercially available pre-etched holes on $Si_3N_4$ membrane (yellow) on $SiO_x$ window (blue). (b-c) AFM scans of bubble curving up (down) and the height profile across the center; the membrane sustains curvatures, which one side of the membrane is compressed, and the other side is elongated to accommodate for the strain gradient (Scale bar: 2 μm).

Here we report a method to engineer a nanodrum geometry to form "bubble" states with bistable strain gradient polarity, which enables versatile writing, reconfiguring, and local control of strain gradients. Figure 1(a) illustrates the schematic representation of these bubble states, which exhibit opposite strain gradient polarities. Atomic force microscopy (AFM) scans (Fig. 1(b) and 1(c)) reveal bistable configurations—up and down states—along with their height profiles. The curvature of the membrane leads to a spatially varying strain along the height, sustaining a stable strain gradient. To investigate and demonstrate mechanical control and design of the bistable bubbles, we use freestanding $La_{0.7}Sr_{0.3}MnO_3$ (LSMO) membranes as an example material [2,18]. While we showcase here our mode of operation using manganite membranes, we note that this procedure can be universally applied to any material in membrane form.

Dorfmeister et al. [19,20] provided a mechanical model predicting bistable state formation based on intrinsic material properties like Young's modulus $E_y$, Poisson's ratio $\nu$ and intrinsic stress $\sigma_0$, and geometric factors including pore radius $r$, membrane thickness $t$, and resulting height $w_0$ of bistability:

$$\sigma_0 = \sigma_c \left(1 + \frac{16}{35}\frac{w_0^2}{t^2}\right), \qquad \sigma_c = \frac{4}{3}\frac{t^2}{r^2}\frac{E_y}{1-\nu^2}$$

When the intrinsic stress $\sigma_0$ is smaller than the critical stress $\sigma_c$, the membrane remains flat over the pore and $w_0 = 0$. When the intrinsic stress $\sigma_0$ of the membrane exceeds the critical stress $\sigma_c$, the rest of the strain energy spontaneously converts to forming bistable strain gradient states with height $w_0$:

$$w_0 = \begin{cases} 0, & \sigma_0 \le \sigma_c \\ \pm\frac{\sqrt{35}}{4}\sqrt{\frac{\sigma_0}{\sigma_c} - 1}, & \sigma_0 > \sigma_c \end{cases}$$

Our experimental observations align qualitatively with these predictions. We analyzed the bistable features across multiple fabricated devices and found a clear correlation between bubble height $w_0$ and membrane thickness. Figure 2(a) presents the statistics of bubble heights for different membrane thicknesses, while the pore diameter is fixed. Error bars represent the statistical standard deviation for $n = 10$-20 bubble profiles on the same sample. Figures 2(b–e) show AFM topography of bubbles formed on holes with diameter of 2 μm, for membrane thicknesses of 20 nm, 30 nm, 40 nm, and 50 nm. The maximum strain difference ($\Delta\varepsilon$) between the top and bottom surfaces of the membrane is inversely proportional to the radius of curvature ($R$). The values of $\Delta\varepsilon$ at apex or valley of the bubble can be estimated geometrically using the following relationship:

$$R = \frac{r^2 + w_0^2}{2w_0}, \qquad \Delta\varepsilon(\%) \approx \pm\frac{t}{2R}$$

When normalized by thickness, one can acquire an estimate of the strain gradient, roughly proportional to $1/R$. Given a fixed pore size, thinner membranes exhibit larger bubble heights $w_0$ due to their lower critical stress $\sigma_c$. As membrane thickness increases, the maximum achievable strain gradient decreases as critical stress $\sigma_c$ (proportional to $t$) increases. Thicker membranes require larger intrinsic stress to induce spontaneous bistability.

Figures 2(f–j) examine the effect of pore radius on bistable bubble height for a fixed membrane thickness of 30 nm. Each sample is fabricated with membranes grown from the same batch on a 5 mm by 5 mm substrate and diced into 9 pieces. $w_0$ shows a tendency to scale with pore size, reaching hundreds of nanometers for pores of 10 μm diameter. Error bars represent the statistical standard deviation for $n$ = 10-20 bubble profiles on the same sample. As discussed above, we anticipate that larger pores would result in the decrease in $\sigma_c$ (inversely proportional to $r^2$) and therefore facilitate enlargement of $w_0$. Below a critical hole radius $r$, no buckling behavior is observed, and the membrane remains flat over the pores. However, when converted to percentage strain variation, larger lateral scales lead to a reduced maximum strain gradient at the bubble apex. This suggests a trade-off between lateral feature size and the strain gradient. In other words, to maximize strain gradient, a thinner membrane on a smaller pore gives more localized curvature with more enhanced bending. As $r/t$ continues to increases, as in Figure 2(j), we observe edge wrinkling behavior as an result of the mechanical instability under axial compression with boundary constraints [21].

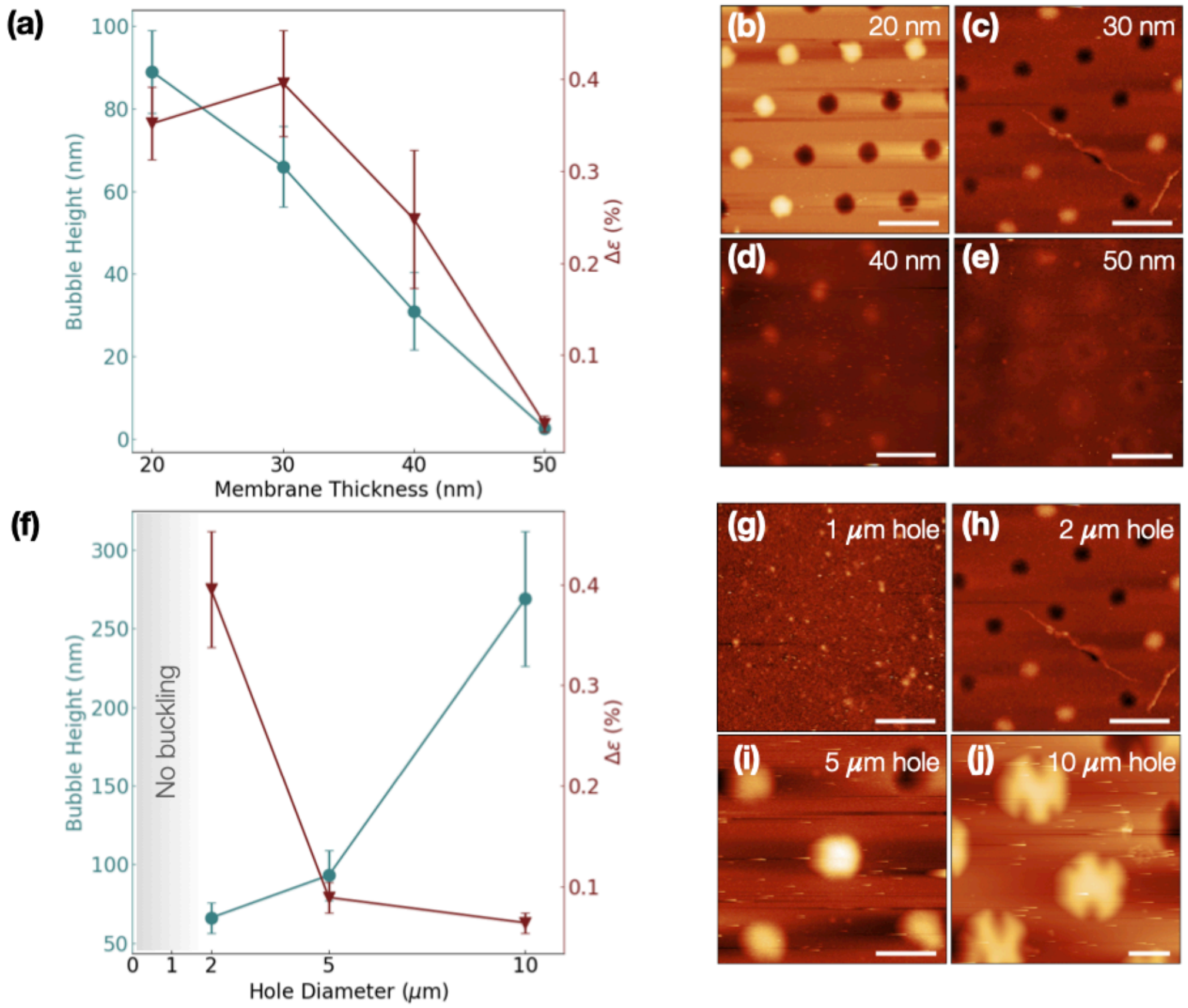

**Figure 2. Statistics of bistable bubble height.** (a) Statistics of bistable bubble height on 2 μm holes collected across different membrane thickness, plotted along with their respective average Δε (%). (b-e) Topography of bubble formations on 2 μm holes for membrane thickness of (b) 20 nm, (c) 30 nm, (d) 40 nm and (e) 50 nm (Scale bar: 5 μm). (f) Statistics of bistable bubble height of 30 nm membrane collected across different hole diameters, plotted along with their respective average Δε (%). (g-j) Topography of bubble formations of 30 nm membrane for hole diameters of (g) 1 μm, (h) 2 μm, (i) 5 μm and (j) 10 μm (Scale bar: 5 μm).

Next, we introduce methods of fabricating bistable strain gradient devices and techniques for programming the bubble morphology through controlling the energy hierarchy between the up and down states. Figure 3(a) is an illustration of the fabrication process. A $CaSr_2Al_2O_6$ sacrificial layer was grown on a $SrTiO_3$ (STO) (001) substrate, followed by the growth of capping layers STO (2 unit cells; u.c.) and LSMO (varying thickness) by pulsed laser deposition (PLD) [2,18]. For most single crystalline thin-film growth modalities, there exist non-zero lattice mismatch between the film and substrate of choice. The intrinsic epitaxial strain can eventually act as a "pre-strain" to the membrane [22]. The solid-solution family of $(Ba,Sr,Ca)_3Al_2O_6$ enables the continuous tuning of the sacrificial layer lattice constant from $a$ = 15.28 Å (4 × 3.82 Å) (Ca 100%) to $a$ = 16.52 Å (4 × 4.13 Å) (Ba 100%) [2,5]. This enables tuning the magnitude of epitaxial pre-strain. The lattice constant of $CaSr_2Al_2O_6$ is at 4 × 3.907 Å, almost perfectly matching with STO substrate lattice constant at $a_{STO}$ = 3.904 Å, and is ~0.7% mismatched to the pseudo-cubic LSMO lattice constant at $a_{LSMO}$ = 3.876 Å. With moderate epitaxial mismatch, the membranes will be strained to match the substrate, maintaining a non-zero pre-strain. Figure 3(b) shows the θ-2θ x-ray diffraction patterns (gray) of STO(4 u.c.)/LSMO(20 nm)/STO(4 u.c.)/$CaSr_2Al_2O_6$(50 u.c.)/STO(001) heterostructure as grown. The few u.c. of STO served as protective capping layers for the LSMO layer, and the LSMO layer is strained to the STO substrate. The inset (left) is the AFM image (scanning area 3 × 3 μm$^2$, scale bar = 1 μm, RMS roughness = 232.8 pm) taken on the step-terraced surface of the sample. The inset (right) is the rocking curve measurement of LSMO (20 nm)/STO (blue, FWHM = 0.0935°) and the STO capped LSMO (20 nm)/sacrificial layer/STO (orange, FWHM = 0.0932°)) as grown, indicating high crystallinity.

The pre-strain is clamped by a commercially available PMMA 950A4 polymer layer, spin-coated at 4000 rpm for 50 seconds and cured at 135°C to assist the release and transfer process.

When released from the original substrate via epitaxial lift-off [5] and transferred onto commercially available (Norcada Inc.) 200 nm thick $Si_3N_4$ with arrays of pores of various diameter (0.5, 1, 2, 5, 10 μm) [9], PMMA holds the membrane at a non-zero residual strain. After the polymer layer is removed by acetone, suspended regions of the membrane undergo compressive restoring force to relax, spontaneously forming bistable bubbles. These bubbles manifest upward or downward curvatures, thus realizing bistable strain gradient states with opposite polarities. The reversible switching of the bistable states will be demonstrated below.

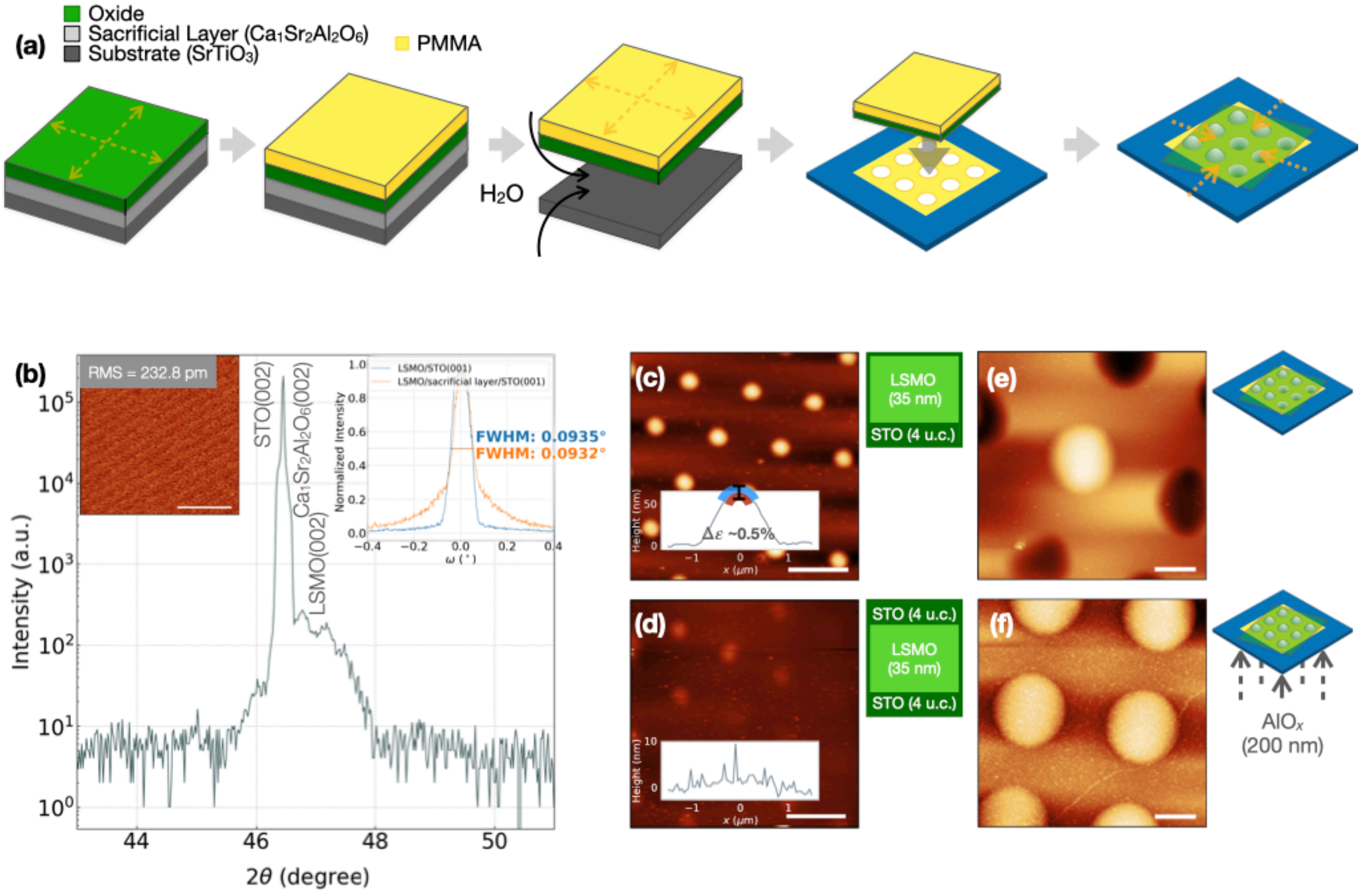

**Figure 3. Fabrication of the bubble morphology and programming of bistable energetics.** (a) Illustration of the fabrication process. In standard epitaxial thin film growth, oxide (green)

lattice is intrinsically pre-strained to the substrate lattice. The pre-strain is clamped by the commercially available PMMA 950A4 polymer layer, spin-coated at 4000 rpm for 50 seconds and cured at 135°C to assist the release and transfer process. Then the membrane is lifted off in water and transfer onto a $Si_3N_4$ porous grid. After removal of PMMA, the oxide membrane undergoes compressive restoring force to relax, spontaneously forming bistable bubbles. (b) θ-2θ x-ray diffraction patterns (gray) of STO(4 u.c.)/LSMO(20 nm)/STO(4 u.c.)/$CaSr_2Al_2O_6$(50 u.c.)/STO(001) heterostructure as grown. The inset (left) is the AFM topographic image (scale bar = 1 μm). The inset (right) is the rocking curve measurement of LSMO (20 nm)/STO (blue, FWHM = 0.0935°) and LSMO (20 nm)/sacrificial layer/STO (orange, FWHM = 0.0932°)) as grown, indicating high crystallinity. (c) Topography of bistable bubble formation with asymmetrically capped bilayer structures. The inset shows the height profile of a bubble. The average $\Delta\varepsilon$ reaches 0.5%. (d) Topography of bistable bubble formation with a symmetrically capped trilayer. The inset shows the height profile which shows a significant suppression of bubble height compared to the asymmetrically capped geometry. (e) Topography of spontaneous formation of bistable bubbles before back filling. (f) Topography of bubbles after back-filling of 200 nm thick $AlO_x$ via e-beam evaporation. All bubbles curve up and become fully supported hemispherical dots. (Scale bars: 5 μm.)

Going further, by designing symmetric and anti-symmetric heterostructures in growth, we can control the relative initial population of the two states and design the energetic anisotropy in downward-switching versus upward-switching. As indicated in discussions above, the larger the ratio of $\sigma_0/\sigma_c$, the more energy is required to switch between the bistable states. Hence, to design the energy hierarchy between the bistable states and to create asymmetry in the switching, one can design the membrane as a heterostructure stack, leveraging intrinsic stress induced from epitaxial mismatch between the constituent layers [22]. For instance, an asymmetrically capped LSMO membrane with STO cap only on one side (Fig. 3(c)) has a non-uniform distribution of stress along the *c*-direction, originating from the ~0.7 % lattice mismatch between LSMO and STO. As a result, the bubbles preferentially stabilized in the same state. On the other hand, with LSMO symmetrically sandwiched between layers of STO as in Figure 3(d), bubble formation is significantly suppressed due to the uniform net strain.

Another means of programming the bubble morphology is through post-synthesis processing. By backfilling the membrane on the pore sites, the strain gradient can be fixed in place, providing a mechanically secure local strain gradient. Figure 3(e-f) demonstrates that through back-filling with $AlO_x$ by e-beam metal evaporation, one can stabilize the bubbles with initially random distributions (Fig. 3(e)) into permanently up-curved and even hemispherical domes (Fig. 3(f)). The samples were flipped to attached to the metal evaporator plate. $Al_2O_3$ target was used to deposit around 100 nm thick of $Al_2O_3$ at deposition rate of 0.3Å/s.

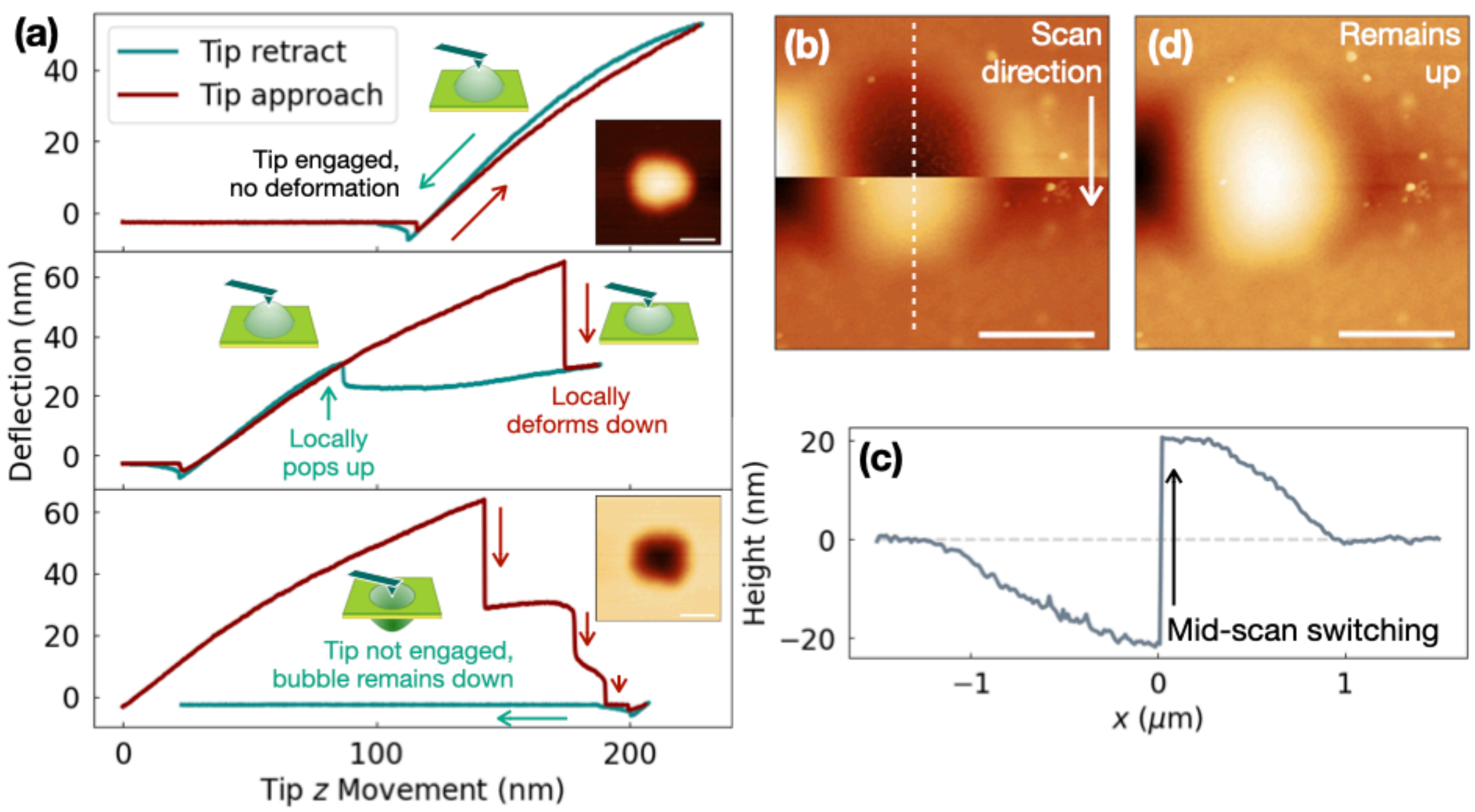

**Figure 4. Switching of the bistable states.** (a) Force-deflection curves illustrate evolution of switching the bistable states, from up to down state, by pressing a 35 nm membrane on 2 μm hole with an AFM tip. Top: No switching occurs; Center: bubble dents during tip approach but later pops back up during tip retraction; Bottom: bubble switched to down state during tip approach (loading) and remains down after tip retraction (unloading). (b) AFM scan of mid-scan switching from down to up state due to tip attraction. (c) Height profile of mid-scan switching. The sudden jump indicates the moment when the bubble pops from the down to the up state. (d) AFM scan showing the bubble remains in the up-state after switching and tip unloading. (Scale bars: 2 μm.)

Now we move on to demonstrate active switching. In Fig. 4, we demonstrate the dynamic control of bistable states via tip manipulation with an AFM in contact mode on a sample of 35 nm LSMO membrane over 2 μm holes. We used a diamond-like-carbon coated force modulation AFM tip (NanoAndMore Inc.) with a nominal tip head of <15 nm in radius with spring constant of 3 N/m. In Fig. 4(a), we demonstrate how the bubble is switched from an up-state to a down-state, confirmed by AFM scans and height profiles. The tip $z$ movement (nm) represents the vertical distance the tip has moved to approach the surface. When the tip is not yet in contact with the surface, the force-deflection curve remains flat – see the flat traces within $0 < z < 120$ nm in the top panel of Fig. 3(a). When the tip touches and presses onto the surface, it begins to exert a loading force onto the surface and the surface responses to deflect the tip. The tip loading force can be accurately obtained by the feedback loop of the AFM. By measuring the force-deflection feedback, we can probe the modes of tip-sample interaction, like bending, stretching, or other surface morphing modes [9,10].

The switching process progresses through three distinct stages, as shown in the evolution of force-deflection curves oat each stage. In the first stage, the loading force monotonically increases with continuous downward movement of the tip but below the threshold to induce switching (Fig. 4(a), top panel). In this regime, the surface reacts as a rigid dome – the tip deflects as it is pressed more onto the surface in a roughly linear relation, while the bubble remains robustly in its up-state – see trend at $z > 120$ nm in the top panel of Fig. 4(a). With further increase of the loading force, the bubble starts to deform but eventually returns to its original state upon tip retraction. The tip remains engaged with the surface and is being pulled down (pushed up) when the bubble surface locally deforms down (up). This results in the sudden dip and jump in the force-deflection curve – see the step-like dip during approaching (jump during retracting) around $z = 180$ nm ($z = 90$ nm) in the middle panel of Fig. 4(a). Finally, in the third stage (Fig. 4(a), bottom panel), sufficiently large force is applied to induce a stable switching of the bubble from the up-state to the down-state. When tip retracts, the bubble remains in the down-state and is no longer in contact with the tip, hence the force-deflection curve remains flat during retraction – see the blue retract trace across the entire range of $z$ in the bottom panel of Fig. 4(a).

Finally, we also demonstrate that this switching is dynamic and reversible. Figure 4(b-c) showed an AFM scan and height profile of an example of mid-scan reverse switching, from the down- to up-state. Here, the attractive interaction between the tip and the membrane is sufficient to overcome the energy barrier for switching. The reverse switching is consistently repeatable within a few attempts of tapping onto the bubble surface by lowering the tip to be in contact with the concave bubble surface and raised back up. Thus the tip-surface attraction can induce the switching of the bubbles from down- to up-state. Figure 4(d) shows that the bubble stably remains in the up state after switching from the down state.

In summary, we present a method for generating switchable strain gradients through bistable nanodrum "bubble" geometries in freestanding oxide thin-film membranes. By leveraging mechanical and geometric design, we demonstrate spontaneous formation and reversible switching of bistable states with opposite strain gradient polarities. We also demonstrate methods of mechanically configuring the bubble morphology by controlling the energy hierarchy between the bistable states. Our results not only validate theoretical modeling of membrane mechanics but also offer a dynamic, scalable approach to modulating local strain gradients. With compatibility across a wide range of oxide materials and measurement modalities, the method opens pathways toward integration with functional device architectures, where on-demand control of strain gradients at the nanoscale can unlock new device functionalities [23–26].

AUTHOR INFORMATION

**Corresponding Author**

Tiffany C. Wang (catwang@stanford.edu); Harold Y. Hwang (hyhwang@stanford.edu)

**Present Addresses**

† Max Planck Institute for Solid State Research, Heisenbergstrasse 1, 70569 Stuttgart, Germany

**Author Contribution**

Synthesis and Fabrication : TCW, VH

Data Acquisition: TCW

Modeling: TCW, MH

Funding Acquisition: HYH

Supervision: HYH

Writing: TCW, MH, HYH

**Funding Sources**

T.C.W., V.H., H.Y.H.: US Department of Energy, Office of Basic Energy Sciences, Division of Materials Sciences and Engineering (contract no. DE-AC0276SF00515).

M.H.: Center for Energy Efficient Magnonics, an Energy Frontier Research Center (CEEMag) funded by the U.S. Department of Energy, Office of Science, Basic Energy Sciences at SLAC National Laboratory under contract # DE-AC02-76SF00515.

ACKNOWLEDGEMENTS

We thank Bai Yang Wang, Woojin Kim, and Yijun Yu for fruitful discussions and guidance.

ABBREVIATIONS

**PMMA,** Poly(methyl methacrylate)